\RequirePackage{amsmath}
\documentclass[runningheads]{llncs}
\usepackage{cite}
\usepackage{algorithmic}
\usepackage{textcomp}
\def\BibTeX{{\rm B\kern-.05em{\sc i\kern-.025em b}\kern-.08em
    T\kern-.1667em\lower.7ex\hbox{E}\kern-.125emX}}

\usepackage[T1]{fontenc}
\usepackage{graphicx}
\graphicspath{ {./figures/} }
\usepackage{amssymb}
\usepackage{amsmath}
\usepackage{amsfonts}
\usepackage{hyperref}
\usepackage{multirow}
\usepackage{makecell}
\usepackage{booktabs}
\usepackage{siunitx}
\usepackage{array}

\usepackage[table]{xcolor}

\usepackage{tabularx}
\usepackage{arydshln}
\usepackage{pdflscape}
\usepackage{adjustbox}
\usepackage{float}

\usepackage{appendix}

\begin{document}
\title{\texorpdfstring{Linking Cryptoasset Attribution Tags to Knowledge Graph Entities: \\ An LLM-based Approach}{Linking Cryptoasset Attribution Tags to Knowledge Graph Entities: An LLM-based Approach}}
\titlerunning{Linking Cryptoasset Attribution Tags to Knowledge Graph Entities}
%
\author{Régnier Avice\inst{1}\orcidID{0000-0002-7320-7868} \and
Bernhard Haslhofer\inst{2}\orcidID{0000-0002-0415-4491} \and
Zhidong Li\inst{1}\orcidID{0000-0002-0784-157X} \and
Jianlong Zhou\inst{1}\orcidID{0000-0001-6034-644X}}
\authorrunning{R. Avice et al.}
%
\institute{University of Technology Sydney, Australia\\
\email {regnier.avice@student.uts.edu.au, \{zhidong.li, jianlong.zhou\}@uts.edu.au} \and
Complexity Science Hub, Vienna, Austria\\
\email{haslhofer@csh.ac.at}}
%

%
\maketitle
\begin{abstract}
Attribution tags form the foundation of modern cryptoasset forensics. However, inconsistent or incorrect tags can mislead investigations and even result in false accusations.
To address this issue, we propose a novel computational method based on Large Language Models (LLMs) to link attribution tags with well-defined knowledge graph concepts. We implemented this method in an end-to-end pipeline and conducted experiments showing that
our approach outperforms baseline methods by up to 37.4\% in F1-score across three publicly available attribution tag datasets. By integrating concept filtering and blocking procedures, we generate candidate sets containing five knowledge graph entities, achieving a recall of 93\% without the need for labeled data. Additionally, we demonstrate that local LLM models can achieve F1-scores of 90\%, comparable to remote models which achieve 94\%. We also analyze the cost-performance trade-offs of various LLMs and prompt templates, showing that selecting the most cost-effective configuration can reduce costs by 90\%, with only a 1\% decrease in performance.
Our method not only enhances attribution tag quality but also serves as a blueprint for fostering more reliable forensic evidence.
\end{abstract}


\section{Introduction}
\label{sec:introduction}

Attribution tags, which link pseudo-anonymous cryptoasset addresses to identifying information about real-world entities and services (e.g., cryptoasset exchanges), form the foundation of modern cryptoasset forensics~\cite{Meiklejohn2013}. Over the past decade, a multibillion-dollar industry has emerged, providing blockchain tracing tools for law enforcement investigations. These tools allow users to ``follow the money'', ultimately leading to the identification and potential conviction of perpetrators. As cryptoassets have gained increasing relevance across various crime sectors, these tools are now used in investigations related to ransomware~\cite{Huang2018, PaquetCloustonRansomware2019}, sextortion~\cite{PaquetCloustonSextortion2019}, and malware~\cite{Gomez2022}. Attribution tags have also been employed to train machine learning models for automatically categorizing service providers, such as exchanges~\cite{Harlev2018, Gomez2022, Liu2022, Zhou2022}, miners~\cite{Harlev2018, Liu2022, Zhou2022}, and ICO wallets~\cite{Liu2022, Zhou2022}. Moreover, these models are used to classify addresses or transactions as illicit~\cite{Liu2022, Zhou2022, Wu2022, Li2022, Hu2023, Chen2019, Bartoletti2020}. Consequently, the accuracy of attribution tags becomes a critical success factor in all these application domains.

Imprecise or incorrect tags can mislead investigations, result in inaccurate model predictions, or, in the worst case, lead to false accusations. As a result, the concept of \emph{attribution tag quality} is gaining increased attention, especially as the scientific validity of crypto-tracing techniques is being more frequently questioned and scrutinized~\cite{Change:2023}. As with any digital forensic investigation that forms the basis for legal decisions, the collected evidence must be reliable~\cite{Frwis2020}, since poor data quality can lead to incorrect conclusions. Ensuring high data quality also guarantees that the methodologies used in digital forensics can be consistently tested and verified.

Since attribution tags originate from and are often shared among multiple stakeholders, data consistency becomes a critical aspect of data quality. A key challenge, for instance, is the inconsistent referencing of real-world entities across different parties~\cite{Frwis2020,Gomez2022}. For example, one party might refer to a specific exchange as \emph{btc-e}, while another uses \emph{btc-e.com}. While a human can easily recognize that both tags refer to the same entity, different tracing tools may interpret them as referring to two distinct entities. To harmonize their representation, reuse, and interpretation in forensic investigations or machine learning tasks, a suitable data format and a computational approach are needed to automatically eliminate potential data inconsistencies.

Given the semantic nature of this practical problem, knowledge graphs offer an effective solution. They represent entities as identifiable semantic concepts rather than plain strings and are widely used in search engines~\cite{GoogleKG2012} and large knowledge bases~\cite{vrandevcic2014wikidata}. The process of linking data to these well-defined entities is known as record linkage~\cite{Elmagarmid2007} or entity linking~\cite{Shen2015}. However, existing approaches often rely on heuristics or heavily labeled, domain-specific training datasets, which are not readily available in the context of cryptoasset attribution tags.

Therefore, in this paper, we present a novel computational approach based on Large Language Models (LLMs) that enables the linking of attribution tag datasets to well-defined knowledge graph concepts without requiring domain-specific fine-tuning. We implemented this approach by integrating several cutting-edge techniques --- such as filtering, blocking, and LLMs --- into an end-to-end pipeline. Our experiments, conducted on three datasets, demonstrate that:

\begin{enumerate}

    \item Our end-to-end entity linking approach outperforms baseline methods across all three datasets, achieving up to a 37.4\% improvement in F1-scores.

    \item $\text{BM25}_3$ blocking and related-concept filtering reduce the number of candidates per tag to 5, with a recall of 93\%, without requiring labeled data.

    \item GPT-4o-based candidate selection achieves a 94\% F1-score, while Mistral 7B-Instruct, which can run locally, achieves a 90\% F1-score.

    \item Using the most cost-effective prompt template reduces costs by 90\%, with only a 1\% drop in performance.

\end{enumerate}

Our method not only enhances the quality of attribution tags but also seeks to inspire broader efforts toward improving data quality, ensuring accurate and reliable evidence in forensic investigations. To ensure reproducibility, we have made our code and datasets publicly available in the following GitHub repository: \url{https://github.com/ravice234/cryptoasset-attribution-tag-linker}.

\section{Background}
\label{sec:background}

\subsection{Attribution Tags}

\begin{figure}
    \centering
    \includegraphics[width=1\textwidth]{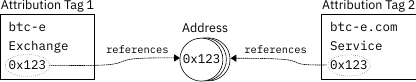}
    \caption{\textbf{Attribution Tag Example}. Two attribution tags referencing the same cryptoasset address \emph{0x123} owned by the real-world entity \emph{BTC-e}.}
    \label{fig:tag-example}
\end{figure}

Attribution tags link pseudo-anonymous blockchain objects, such as addresses or transactions, to real-world actors or events. They provide additional context, such as the name of a service controlling an address (e.g., btc-e), some form of categorization (e.g., exchange), and any other information that might be useful in forensic investigations. Figure~\ref{fig:tag-example} illustrates an example in which two distinct attribution tags originating from different sources reference the same cryptoasset address \texttt{0x123} and describe the same real-world actor, a well-known cryptoasset exchange. One can observe that, despite describing the same entity, the attribution tag data records the name (\texttt{btc-e} vs. \texttt{btc-e.com}) and categorize the exchange differently (\texttt{Exchange} vs. \texttt{Service}).

Attribution tag data quality issues can occur at various levels~\cite{Haslhofer2010}: technical heterogeneities, such as different data formats, can impede uniform processing; syntactic heterogeneities, like the use of different encoding schemes, can hinder uniform interpretation; and semantic heterogeneities, such as the use of different names to denote the same real-world concept (synonyms, homonyms, hypernyms, etc.), can lead to inconsistent interpretations. In this paper, we primarily focus on resolving semantic interoperability issues, assuming that technical and syntactic issues can be addressed using common data preprocessing procedures.

Introducing knowledge graphs to data management environments has become a common strategy to deal with semantic interoperability issues. A knowledge graph defines nodes representing real-world entities of interest and semantic relationships between these entities~\cite{Hogan2021}. An example is eBay's product knowledge graph, which allows them to identify if two sellers sell the same products, or if the products are related otherwise. 

\begin{figure}
    \centering
    \includegraphics[width=0.9\textwidth]{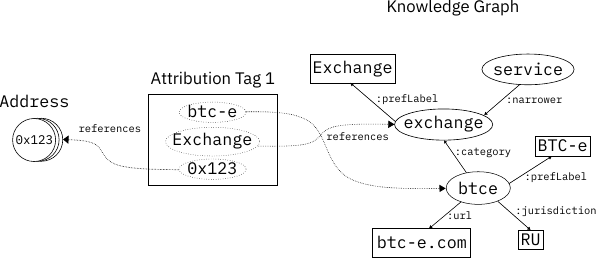}
    \caption{\textbf{Linking an Attribution Tag to the Knowledge Graph}. Attribution tag instances are linked to concepts defined in the knowledge graph.}
    \label{fig:knowledge-graph}
\end{figure}

The process of linking text mentions to entities in a knowledge graph is called \emph{entity linking} \cite{Shen2015}. This process can be further broken down into: 1) generating a subset of entities (\emph{candidates}) that are most likely to match, and 2) selecting which, if any, candidate matches to the text mention. Candidate set generation can be achieved by excluding implausible (\emph{filtering}) and clustering similar (\emph{blocking}) entities \cite{Papadakis2020}. The selection of a matching candidate is typically performed using a decision function learned through machine learning. In this paper, the term \emph{entity linking} is used to describe the process of linking tags to a knowledge graph; while our solution also encompasses techniques that stem from the closely related problem of linking database records (record linkage, entity matching), for the sake of simplicity, we will consistently refer to it as entity linking.

\subsection{Related Work}

Entity linking approaches where tuples are linked to knowledge base entities have utilized look-up methods \cite{Ritze2015}, embedding comparisons \cite{Deng2020}, and hybrid approaches \cite{Efthymiou2017}. Record-based entity linking has seen advancements using classical machine learning \cite{Konda2016}, deep neural networks \cite{Mudgal2018}, and pre-trained language models \cite{Li2020}. In \cite{Tu2023}, a mixture-of-experts approach was proposed, utilizing the training results from various data integration matching tasks. For blocking, many state-of-the-art approaches use deep learning \cite{Li2020, Thirumuruganathan2021, Wang2023, Brinkmann2024}. A simple tf-idf based blocker can achieve competitive results without training and labeled data as shown in \cite{Paulsen2023}.

More recently, researchers started to examine LLMs on data integration tasks. Studies have evaluated and proposed various models, including GPT-3 \cite{Narayan2022}, Jellyfish variants \cite{Zhang2024}, and others \cite{Peeters2024}, on tasks such as entity linkage, data imputation, and error detection. Different matching strategies for LLMs in detail have also been explored in \cite{Peeters2024, Wang2024}.

Within the specific domain of cryptoasset investigations, \cite{Gomez2022} described a method for resolving conflicting attribution tags. They use the edit distance to harmonize strings referring to the same entity. The application of LLMs in the context of cryptoassets and blockchains has been explored for multiple tasks, such as detecting anomalous Ethereum transactions \cite{gai_blockchain_2023}, auditing smart contracts \cite{david_you_2023}, and identifying discrepancies between smart contract bytecode and project documentation \cite{gan_defialigner_2024}.

In this paper, we go beyond this approach, by linking attribution tag datasets to well-defined knowledge graph concepts using an LLM-based approach. 

\section{Data}
\label{sec:dataset}

In this section, we present the data that informed the design and implementation of our approach and was used throughout our experiments.

\subsection{GraphSense TagPacks}
\label{subsec:data_gs}

GraphSense is an open-source cryptoasset analytics platform that provides a curated collection of over 500,000 publicly available attribution tags. A TagPack is a data structure used to package and share attribution tags~\footnote{https://github.com/graphsense/graphsense-tagpacks}. Each tag corresponds to a blockhain address and includes a label, several optional fields, and categorization information. The categories are drawn from a subset of the \emph{INTERPOL Dark Web and Virtual Assets Taxonomy (DWVA)}\footnote{https://misp-galaxy.org/interpol-dwva/}, a community-driven effort to define common forms of abuse and entities representing real-world actors and services within the broader Darknet and Cryptoasset ecosystems.

In addition, GraphSense provides a curated list of 2,862 actors, each representing a well-defined real-world entity within the cryptoasset ecosystem. These actors encompass a wide range of roles and types in the industry such as centralized exchanges (e.g., Binance), decentralized finance platforms (e.g., Aave), and mixing services (e.g., Tornado Cash). Each actor is assigned a unique ID, a label, one or more categories from the DWVA taxonomy, and optional fields such as a URL or jurisdiction. This enables GraphSense to partially implement a knowledge graph, offering explicit links between entities, which we use as a ground-truth dataset. In total, 378,550 attribution tags contain such actor links.

Since many attribution tags share the same labels and actor links and we are only interested in unique records, we filtered out duplicates, leaving 2,570 unique linked attribution tags. These were split into training, validation, and test datasets in a 1:30:69 ratio. The training set was used to create few-shot examples, while the validation set was employed in experiments 1 and 2 to optimize the individual components. The test set was exclusively used in experiment 3 to evaluate our approach end-to-end.

\subsection{WatchYourBack Attribution Tags}
\label{subsec:data_wyb}

We utilize the dataset published in \cite{Gomez2022}. Structurally, these attribution tags are similar to the GraphSense TagPack, as they also include categories and subcategories. However, their taxonomy is not harmonized with the GraphSense Actor Taxonomy.

The dataset consists of tags aggregated from various sources, including GraphSense and other shared datasets. To avoid duplicating tags, we filter out any that are linked to addresses already present in the GraphSense TagPack database. From this filtered set, we manually selected and annotated 126 records, 67 of which contain an actor link.

\subsection{DeFi Rekt Database}
\label{subsec:data_defirekt}

The DeFi Rekt database \cite{Defirekt2024} contains over 3,500 events related to crimes involving cryptoassets. Many of these events involve actors within the ecosystem. Each event includes a title and optional fields such as the date of the event, and funds involved. For our experiments, we randomly sampled 100 records from events where the loss of funds exceeded 100,000 USD. We manually annotated these sampled events with actor links. Out of the 100 sampled records, only 32 contained an actor link.

\subsection{Dataset summary}
\label{subsec:data_summary}

Table~\ref{tab:datasets} summarizes the data we used for design and implementation of our approach and the experiments we conducted. The total number of distinct actors is lower than the sum of individual datasets due to overlapping actors between datasets.

\begin{table}
    \centering
    \caption{Overview of the three attribution tag datasets used in this study.}
    \begin{tabular*}{\textwidth}{@{\extracolsep{\fill}}lrrr}
    \toprule
    \textbf{Dataset} & \textbf{Samples} & \textbf{Actor Links} & \textbf{Distinct Actors}\\
    \midrule
    GraphSense Tag Pack & 2,570 & 2,570  & 484\\
    \quad \color{gray}{Train} & \color{gray}{25} & \color{gray}{25} & \color{gray}{20}\\
    \quad \color{gray}{Validation} & \color{gray}{771} & \color{gray}{771} & \color{gray}{198}\\
    \quad \color{gray}{Test} & \color{gray}{1,774} & \color{gray}{1,774} & \color{gray}{361}\\
    WatchYourBack & 126 & 67 & 58\\
    Defi Rekt & 100 & 32 & 32\\
    \midrule
    Total & 2,794 & 2,669 & 520\\
    \bottomrule
\end{tabular*}

    \label{tab:datasets}
\end{table}


\section{Approach}
\label{sec:approach}

\begin{figure}
    \centering
    \includegraphics[width=\textwidth]{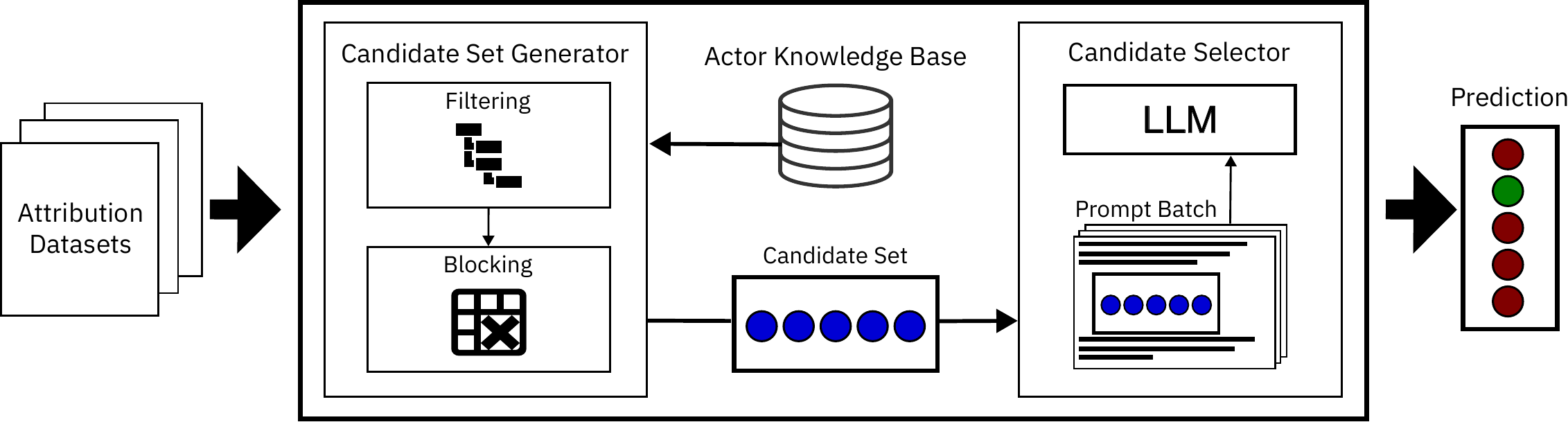}
    \caption{\textbf{Approach Overview}. The \emph{candidate set generator} filters potential entities and the \emph{candidate selector module} identifies the matching entity.}
    \label{fig:overview}
\end{figure}

Our approach for linking attribution tags to knowledge graph concepts comprises two main components, as illustrated in Figure~\ref{fig:overview}: the \emph{candidate set generator} and the \emph{candidate selector} modules. The candidate set generator reduces the pool of potential actors in the knowledge graph that may correspond to an attribution tag by applying filtering and blocking techniques. Next, the candidate selector module determines which, if any, of the proposed entities match the attribution tag. The following sections offer a detailed explanation of each module.

\subsection{Candidate Set Generator}
\label{subsec:csg}

The goal of the candidate set generator is to reduce the pool of candidates in the knowledge graph that potentially match a given attribution tag. Since LLM inference is expensive and pairwise comparison is of order $O(nm)$, where $n$ is the number of records and $m$ is the number of entities in the knowledge graph, an approach that minimizes the comparisons is required.

To reduce the overall cost, we limit the entities in each prompt to $k$ candidates by applying \emph{filtering} and \emph{blocking} techniques, thereby reducing the problem complexity to $O(nk)$, where $k \ll m$. Filtering is the process of eliminating incorrect candidates, while blocking refers to similarity-based clustering that identifies likely and unlikely candidates \cite{Papadakis2020}.

\subsubsection{Filtering}

The candidate set can be narrowed down when both the attribution tag and the corresponding knowledge graph entity are associated with categorization information from the same controlled vocabulary. For example, an attribution tag might be categorized as ``exchange'' and a knowledge graph entity as ``service'', where the latter represents a semantically broader concept than the former. When such information is available, as in the GraphSense TagPack dataset (see Section~\ref{subsec:data_gs}), it can be leveraged for filtering. We define two possible filtering methods:

\begin{itemize}
    
    \item \emph{Same-Concept Filtering}: This method excludes actors that belong to a category different from the one specified in the attribution tag.

    \item \emph{Related-Concept Filtering}: This approach is more flexible than same-concept filtering, as it leverages the taxonomy structure to exclude all actors associated with concepts unrelated to the attribution tag. Related concepts include both ancestors and descendants of the original concept, but exclude descendants of ancestor concepts that are not directly related.

\end{itemize}

\subsubsection{Blocking}

The basic idea behind blocking is to avoid comparing entities that are unlikely to match, significantly reducing the number of comparisons. This is done by partitioning the data into smaller subsets (\emph{blocks}), where entities share some similarities or common attributes.  Inspired by \cite{Paulsen2023}, who demonstrate that simple blocking methods --- without requiring machine learning or pre-trained models --- can achieve strong results, we apply two straightforward methods:

\paragraph{$\text{BM25}_3$}
This method is based on the Okapi BM25 \cite{Robertson1994} that is part of the family of term frequency-inverse document frequency (tf-idf) scoring functions. We tokenize the document and query strings into trigrams as proposed in \cite{Brinkmann2024} to allow for approximate string matching. The overall score for a tokenized document \( D \) with respect to a tokenized query \( Q \) is given by:

\begin{equation}
    \text{BM25}(D, Q) = \sum_{i=1}^{n} \text{idf}(q_i) \cdot \frac{f(q_i, D) \cdot (k_1 + 1)}{f(q_i, D) + k_1 \cdot (1 - b + b \cdot \frac{|D|}{\text{avgdl}})}
\end{equation}

where \( f(q_i, D) \) is the frequency of token \( q_i \) of query \( q\) in document \( D \) and \( \text{avgdl} \) is the average document length in the corpus. The inverse document frequency (IDF) is calculated as:

\begin{equation}
    \text{idf}(q_i) = \log \left( \frac{N - n(q_i) + 0.5}{n(q_i) + 0.5} + 1 \right)
\end{equation}

We use the standard parameters \( k_1 = 1.5 \) and \( b = 0.75 \) defined in the implentation of the rank\_bm25 library\footnote{https://github.com/dorianbrown/rank\_bm25}.

\paragraph*{$\text{Overlap}_3$}
This method measures the similarity between two strings based on the overlap $\frac{|A \cap B|}{min(|A|,|B|)}$ of their trigram sets $A$ and $B$.

\subsection{Candidate Selector}
\label{subsec:cs}

The candidate selector module takes a set of candidates for each attribution tag and selects the best matching entity. Technically, this step is implemented using an LLM. In the first stage, the module constructs a batch of prompts, where each prompt corresponds to an attribution tag and includes all associated candidates. Optionally, the prompts can include examples (few-shot prompts) showcasing both matching and non-matching cases. This prompt batch is then fed to the LLM, which is tasked with either selecting the candidate that best matches the attribution tag or indicating that none of the candidates correspond to the tag.

\begin{figure}
    \centering
    \includegraphics[width=0.95\textwidth]{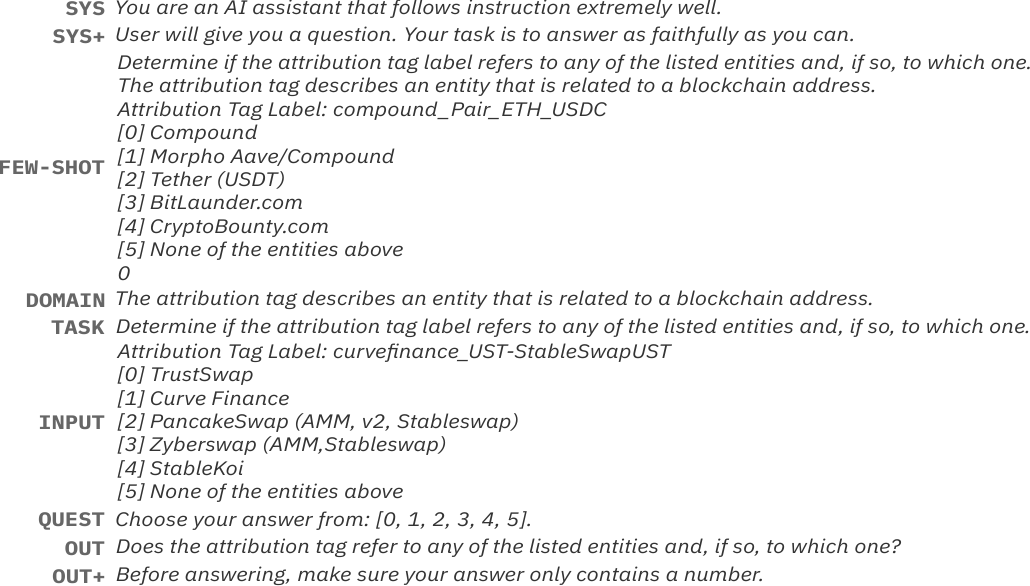}
    \caption{\textbf{Prompt template}. The template used for prompting candidate selection. It consists of several parts (e.g., SYS, FEW-SHOT, INPUT, etc.) and defines the instructions provided to an LLM to guide its response or generated output.}
    \label{fig:el_base_prompt}
\end{figure}

Figure~\ref{fig:el_base_prompt} shows the template used for this prompting task. It consists of an extended system message (SYS, SYS+), the few-shot examples (FEW-SHOT), a task description (TASK), a domain statement (DOMAIN), the input data (INPUT), the selection question (QUEST), and an extended output format reminder (OUT, OUT+).

\section{Experiments}
\label{sec:experiments}

\subsection{Setup}

\paragraph{Models} The candidate selector module (see Section~\ref{subsec:cs}) leverages LLMs, integrating both remote models via the OpenAI API and locally hosted models. For our experiments, we utilized two remote and six local models, all compatible with consumer-grade hardware:

\begin{itemize}

    \item \emph{GPT-4o, GPT-3.5 Turbo}: These models, hosted by OpenAI, were employed in their specific versions: gpt-3.5-turbo-0125 and gpt-4o-2024-05-13.

    \item \emph{Jellyfish-7B, Jellyfish-13B-AWQ}: Developed and fine-tuned for data preprocessing tasks \cite{Zhang2024}, the Jellyfish models include a 7B version, based on Mistral-7B-Instruct-v0.2, and a 13B version, based on OpenOrca-Platypus2-13B. The 13B model was quantized to fit on consumer-grade GPUs using Activation-aware Weight Quantization (AWQ) \cite{Lin2024}.

    \item \emph{Mistral 7B/7B-Instruct}: Both the base and instruction fine-tuned versions of the Mistral 7B model, developed by Mistral AI, were evaluated in their v0.3 release.

    \item \emph{Meta LLama 3 8B / 8B-Instruct}: These models, developed by Meta, include both the base and instruction fine-tuned versions of the Meta LLama 3 8B model.

\end{itemize}

\paragraph{Prompt Configurations} Prompt engineering can significantly influence LLM performance in tasks such as entity linking, as demonstrated by \cite{Peeters2024}. The authors show that different template structures yield better results with certain models, while other templates are more effective with different models. Building upon the structure of our base prompt template (see Figure~\ref{fig:el_base_prompt}), we are testing nine additional shorter template confirmations, as detailled in Table~\ref{tab:el_template_config}. In all ten configurations, we are considering few-shot examples (FEW-SHOT) as optional and place them before the task description (TASK).

\begin{table}
    \centering
    \caption{\textbf{Prompt template configurations}. Each template represents a variation of the structural elements from our base prompt template in Figure~\ref{fig:el_base_prompt}.}
    \begin{tabular*}{\textwidth}{@{\extracolsep{\fill}}ll}
     \toprule
     \textbf{Template} & \textbf{Configuration}  \\
     \midrule
     0 & SYS, SYS+, TASK, DOMAIN, QUEST, OUT, OUT+\\
     1 & SYS, TASK, DOMAIN, QUEST, OUT, OUT+\\
     2 & SYS, SYS+, TASK, DOMAIN, QUESTION, OUT\\
     3 & SYS, SYS+, TASK, QUEST, OUT, OUT+\\
     4 & SYS, SYS+, TASK, DOMAIN, QUESTION\\
     5 & TASK, DOMAIN, QUEST, OUT, OUT+\\
     6 & SYS, TASK, QUEST, OUT\\
     7 & TASK, QUESTION, OUT, OUT+\\
     8 & SYS, TASK, QUEST\\
     9 & TASK, QUEST\\
     \bottomrule
\end{tabular*}
    \label{tab:el_template_config}
\end{table}

\paragraph{Hardware} All experiments are conducted on an AWS g5.xlarge instance equipped with 16 GB of RAM, 4 vCPUs powered by a second-generation AMD EPYC processor, and an NVIDIA A10G Tensor Core GPU with 24 GB of VRAM. For all models, we set the temperature to 0. Local models are run using vLLM\cite{Kwon2023}, an LLM serving system, with the \texttt{max\_num\_of\_seq} parameter set to 128 and \texttt{enable\_prefix\_caching} to true.

\subsection{Experiment 1: Candidate Set Generation}
\label{sec:exp1}

The goal of this experiment is to evaluate different filtering and blocking techniques (described in Section~\ref{subsec:csg}) and find an appropriate candidate set size. We run the experiment on the GraphSense Tag Pack validation dataset on candidate set sizes ($k$) of 1, 5, 10, and 25. The task is to predict the correct actor for each sample within the $k$ candidates by comparing the tag label with the actor label, e.g., \texttt{btc-e.com} with \texttt{BTC-e}. To evaluate the performance of combined filtering and blocking techniques, we compute the ratio of attribution tags for which the correct actor link is included among the $k$ candidates. This metric is often referred to as top-k accuracy. However, since each attribution tag has exactly one correct actor link, the proportion of correctly recovered actor links, \emph{recall}, is equivalent in this context. Following prior works \cite{Papadakis2020, Paulsen2023} on blocking methods, we refer to this metric as recall. Note that in this scenario, precision depends solely on $k$ and recall, and thus does not provide additional information.

\begin{table}
    \centering
    \caption{\textbf{Effectiveness of candidate set generation}. Recall of candidate set generation for different candidate set sizes ($k$), blocking ($\text{BM25}_3$, $\text{Overlap}_3$), and filtering (\emph{same-concept}, \emph{related-concept}) techniques.}
    \begin{tabular*}{\textwidth}{@{\extracolsep{\fill}}llcccc}
    \toprule
    \multirow{2}{*}{\textbf{Filter}} & \multirow{2}{*}{\textbf{Blocker}} & \multicolumn{4}{c}{\textbf{Recall}} \\
    \cmidrule(lr){3-6} & & $k$=1 & $k$=5 & $k$=10 & $k$=25 \\ 
    \midrule
    \multirow{2}{*}{No Filtering}    & $\text{Overlap}_3$ & 0.467 & 0.826 & 0.872 & 0.904 \\ 
                                     & $\text{BM25}_3$    & \underline{0.765} & \underline{0.899} & \underline{0.916} & \textbf{0.957} \\
    \addlinespace[0.5em]
    \multirow{2}{*}{same-concept}    & $\text{Overlap}_3$ & 0.553 & 0.739 & 0.763 & 0.774 \\ 
                                     & $\text{BM25}_3$    & 0.669 & 0.770 & 0.776 & 0.782 \\
    \addlinespace[0.5em]
    \multirow{2}{*}{related-concept} & $\text{Overlap}_3$ & 0.658 & 0.873 & 0.914 & 0.938 \\ 
                                     & $\text{BM25}_3$    & \textbf{0.811} & \textbf{0.933} & \textbf{0.944} & \underline{0.955} \\ 
    \bottomrule
\end{tabular*}

    \label{tab:cg_recall_scores}
\end{table}

The results in Table~\ref{tab:cg_recall_scores} demonstrate that $\text{BM25}_3$ consistently outperforms $\text{Overlap}_3$ across all candidate set sizes $k$. Using the \emph{same-concept} filter degrades performance, suggesting that it is overly restrictive. In contrast, the related-concept filter produces slightly better candidate sets than using no filtering. The candidate sets improve only marginally when their size exceeds 5. Therefore, we conclude that for our subsequent experiments, $k=5$ is the optimal candidate set size and $\text{BM25}_3$ is the preferred blocking method. The related-concept filter will be applied where relevant. An illustration of the performance on the different candidate set sizes can be found in \ref{appendix:cg} 

\emph{This demonstrates that basic blocking techniques, such as the $\text{BM25}_3$ blocker, are effective in generating candidate sets with as little as 5 elements, achieving 90\% recall, and that related-concept filtering can further improve recall to 93\%.}


\subsection{Experiment 2: Candidate Selection}
\label{sec:exp2}

The goal of our second experiment is to assess the performance of different LLM-based candidate selectors using various prompt template configurations (see Table~\ref{tab:el_template_config}) and few-shot examples. We run the experiment on the GraphSense TagPack validation set, using the candidate sets generated by the previous experiment. The model's task is to select the correct actor for each attribution tag from the candidate set if present, otherwise to predict that no candidate matches. The model selects a candidate by responding with the corresponding number, as illustrated in Figure~\ref{fig:el_base_prompt}. Invalid responses are classified as no match, following the guidelines of \cite{Narayan2022, Peeters2024}. We treat the task as a multiclass classification problem and evaluate the models using accuracy and macro-averages for recall, precision, and F1-score.

\begin{table}
    \centering
    \caption{Candidate Selector results on the GraphSense TagPack validation set, showing recall ($\textbf{R}$), precision ($\textbf{P}$), F1-score ($\textbf{F1}$), and accuracy ($\textbf{Acc.}$) for each model’s top-performing template ($\textbf{T}$). Best performance is in bold, second best is underlined.}
        \begin{tabular*}{\textwidth}{@{\extracolsep{\fill}}lcccccccccc}
    \toprule
    \multirow{2}{*}{\textbf{Model}} & \multicolumn{5}{c}{\textbf{Zero-Shot}} & \multicolumn{5}{c}{\textbf{Five-Shot}} \\
    \cmidrule(lr){2-6} \cmidrule(lr){7-11}
     & \textbf{T} & \textbf{R} & \textbf{P} & \textbf{F1} & \textbf{Acc.} & 
       \textbf{T} & \textbf{R} & \textbf{P} & \textbf{F1} & \textbf{Acc.} \\
    \midrule
    GPT4o & 1 & \textbf{0.907} & \textbf{0.919} & \textbf{0.910} & \textbf{0.953} & 7 &\textbf{0.939} & \textbf{0.940} & \textbf{0.939} & \textbf{0.983} \\
    GPT3.5 & 7 & 0.536 & 0.573 & 0.535 & 0.499 & 9 &\underline{0.929} & \underline{0.928} & \underline{0.927} & 0.951 \\
    Jellyfish 7B & 0 & \underline{0.804} & \underline{0.802} & \underline{0.797} & 0.847 & 3 & 0.876 & 0.870 & 0.869 & 0.946 \\ 
    Jellyfish 13B & 6 & 0.568 & 0.588 & 0.571 & 0.708 & 2 & 0.797 & 0.790 & 0.789 & 0.879 \\ 
    Llama 3 8B & 0 & 0.744 & 0.743 & 0.738 & 0.827 & 0 & 0.868 & 0.859 & 0.861 & 0.947 \\ 
    Llama 3 8B-Inst & 7 & 0.729 & 0.731 & 0.725 & 0.844 & 3 & 0.883 & 0.874 & 0.875 & 0.955 \\
    Mistral 7B & 1 & 0.737 & 0.732 & 0.728 & \underline{0.850} & 8 & 0.829 & 0.813 & 0.812 & 0.908 \\ 
    Mistral 7B-Inst & 7 & 0.711 & 0.702 & 0.698 & 0.799 & 0 & 0.913 & 0.900 & 0.903 & \underline{0.958} \\ 
    \bottomrule
\end{tabular*}
    \label{tab:cs_performance}
\end{table}

The results in Table~\ref{tab:cs_performance} indicate that GPT-4o outperforms all other methods in both zero-shot and five-shot scenarios. Notably, while GPT-3.5 has the weakest performance with zero examples, it achieves the second-highest F1-score when five examples are included in the prompt. The Jellyfish 7B model ranks second in zero-shot F1-score, suggesting that its fine-tuning on diverse data preprocessing tasks positively impacts the entity selection task. In contrast, the Jellyfish 13B model underperforms compared to the other local models, despite having more parameters, confirming the findings of \cite{Zhang2024} that it performs worse on unseen tasks than its 7B counterpart. Among the local models, Mistral 7B-Instruct performs the best, achieving an F1 score of more than 90\%.

\begin{figure}
    \centering
    \includegraphics[width=\textwidth]{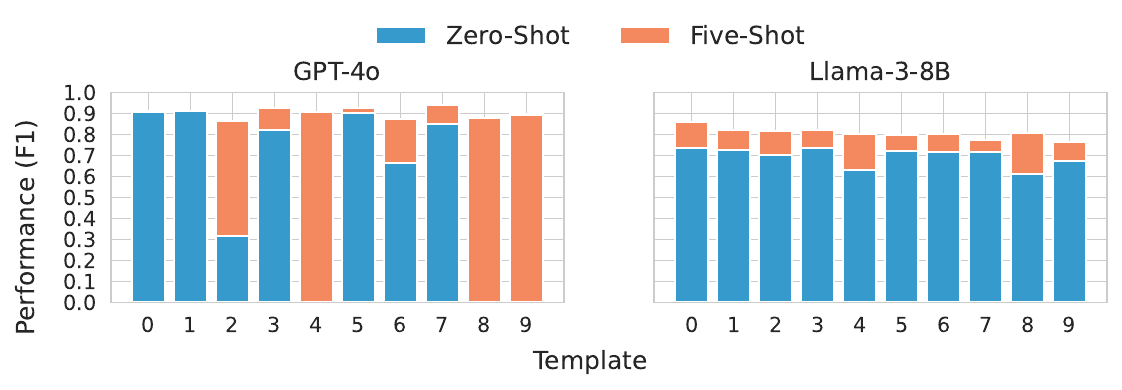}
    \caption{\textbf{Model Performance with different Templates.} \emph{GPT-4o's} zero-shot results vary significantly across different templates, while \emph{Llama-3 8B} has a more stable performance.}
    \label{fig:cs_result}
\end{figure}

In Figure~\ref{fig:cs_result}, we can see that the template choice can have a significant effect on some models. For example, GPT-4o zero-shot results are close to zero on some templates, while on others, they achieve F1-scores of over 90\%. The Meta Llama 3 base model has the most stable performance and is the only model that achieves more than 60\% F1-score on every template with zero examples. An overview of all models can be found in \ref{appendix:el}.

\subsubsection{Cost Analysis}

Choosing more extensive template configurations and introducing few-shot examples increases the size of the prompts, and thus makes inference more expensive. For remote models, we define cost in \$USD based on the OpenAI API usage policy\footnote{https://openai.com/api/pricing/} that charges based on the prompt size. The current rates are (\$5,\$15) per 1 million input/output tokens for GPT-4o and (\$0.5,\$1.5) for GPT-3.5 Turbo. 

For local LLMs, we use inference time as the cost metric. We define inference time as the time taken by the model to process all prompt batches and generate the responses, which includes tokenization but does not include the loading of the weights or any prompt pre/post-processing. We calculate the average run time of five runs for each template in both zero- and five-shot settings. 

To determine which models and what configurations provide the best cost-performance value we first define value as:
\begin{equation}
    V_{\text{\tiny T/S}} = \text{F1}_{\text{\tiny T/S}} * (1 - \Tilde{C}_{\text{\tiny T/S}})
\end{equation}
where {\text{\footnotesize T/S}} is the template/shot configuration and $\Tilde{C} = \frac{C - \min(C)}{\max(C) - \min(C)}$ is the normalized cost. We normalize the cost for remote and local models separately. 

\begin{table}
    \centering
    \caption{Template/Shots (\textbf{T/S}) configuration and cost and performance differences ($\mathbf{\Delta C/F1}$) between models with the highest F1-score and the one with the best cost-performance value ($\mathbf{V}$). Local and remote models are compared separately}
    \begin{tabular*}{\textwidth}{@{\extracolsep{\fill}}lrrrrrr}
\toprule
\multicolumn{1}{c}{\textbf{Model}} & \multicolumn{1}{c}{\textbf{T/S}} & \multicolumn{1}{c}{$\mathbf{F1}$} & \multicolumn{1}{c}{$\mathbf{C}$} & \multicolumn{1}{c}{$\mathbf{V}$} & \multicolumn{2}{c}{$\mathbf{\Delta (\%)}$} \\
& & & & &\multicolumn{1}{c}{$\mathbf{C}$} & \multicolumn{1}{c}{$\mathbf{F1}$} \\
\midrule
GPT-4o  & 7/5 & 0.939 & \$3.359 & 0.116 & & \\
GPT-3.5 & 9/5 & 0.927 & \$0.256 & 0.876 & -92.380 & -1.163 \\
\midrule
Mistral-7B-Inst & 0/5 & 0.903 & 33.951s & 0.659 & & \\
Mistral-7B-Inst & 9/5 & 0.899 & 22.777s & 0.797 & -32.941 & -0.390 \\
\bottomrule
\end{tabular*}
    \label{tab:cs_cost}
\end{table}

In Table \ref{tab:cs_cost} we see that we can save over 90\% on costs when using GPT-3.5 with template 9 while only losing 1\% on the F1-score compared to the best GPT-4o configuration. The template configuration for local LLMs has less cost impact because we can reduce redundant computation to encode the large shared prefix of the prompts by using vLLM prefix caching \cite{Kwon2023}. However, using Mistral-7B-Inst with template 9 instead of template 0 still reduces inference time by more than 30\%, with a performance decrease of less than 0.5\%.

\emph{This experiment shows LLM-based entity linking is effective, with GPT-4o achieving a 94\% F1-score. Aditionally, LLM's that can run locally in a consumer-grade GPU perform well, reaching 90\% F1-score. Furthermore, we show that cost-performance analysis can yield a 90\% cost reduction for only 1\% of performance decrease.}

\subsection{Experiment 3: End to End Entity Linking}

The goal of this experiment is to test our approach of linking attribution tags to a knowledge graph end-to-end and compare it to baseline solutions. We run the experiment on the GraphSense TagPack test set, WatchYourBack, and Defi Rekt datasets. All samples pass through the candidate generator, and the resulting batch of candidate sets are then fed to the candidate selector. For the candidate set generator, we employ the $\text{BM25}_3$ blocker. Additionally, we apply the \emph{related-concept} filtering for the GraphSense dataset, while no filtering is used for the other datasets, because their categories are not linked to knowledge graph concepts. For the candidate selector, we use for each LLM the template that achieved the best performance in the previous experiment. We follow experiment 2's evaluation method, with the difference that a miss by the candidate set generator is counted as an error. We compare the LLMs with the following baseline methods:

\paragraph{$\text{BM25}_3$:} We use the top ranking candidate of our $\text{BM25}_3$ blocker, and decide based on a threshold if it is a match or not. The threshold of $\mathit{15.7238}$ was determined by evaluating the model’s precision and recall on all top-candidate scores in the GraphSense TagPack validation set, and selecting the one that maximizes the F1-score.

\paragraph{UnicornPlus, UnicornPlusFT:} UnicornPlus\cite{Tu2023} is a DeBERTa-based mixture-of-expert model that is fine-tuned for data integration matching tasks. For a fair comparison, we apply the same candidate set generation process and perform pairwise matching between attribution tags and their candidates. In case multiple candidates match, we apply the softmax function on each prediction and choose the one with the highest probability. Furthermore, we create UnicornPlusFT, a fine-tuned version of the model. For this, we reshuffle our GraphSense train and validation sets with a 80/20 training/validation split and train the model for 10 epochs using the settings proposed in \cite{Tu2023}.

\begin{table}
    \centering
    \caption{Performance of various models on different datasets}
    \begin{tabular*}{\textwidth}{@{\extracolsep{\fill}}lccccccc}
\toprule
\multirow{2}{*}{\textbf{Model}} & \multicolumn{2}{c}{\textbf{GraphSense}} & \multicolumn{2}{c}{\textbf{WatchYourBack}} & \multicolumn{2}{c}{\textbf{DeFi Rekt}} \\
\cmidrule(lr){2-3} \cmidrule(lr){4-5} \cmidrule(lr){6-7}
 & \textbf{F1} & \textbf{Acc.} & \textbf{F1} & \textbf{Acc}. & \textbf{F1} & \textbf{Acc.} \\
\midrule
$\text{BM25}_3$             & 0.718 & 0.789 & 0.495 & 0.516 & 0.393 & 0.510 \\
UnicornPlus                 & 0.667 & 0.445 & 0.558 & 0.651 & 0.352 & 0.670 \\
UnicornPlusFT               & 0.783 & 0.873 & 0.542 & 0.603 & 0.419 & 0.610 \\
\\
GPT4o                       & \textbf{0.853} & \textbf{0.927} & \textbf{0.801} & \textbf{0.873} & \textbf{0.793} & \textbf{0.930} \\
GPT3.5                      & 0.810 & 0.881 & \underline{0.756} & \underline{0.825} & \underline{0.691} & \underline{0.890} \\
Llama 3 8B-Inst             & 0.814 & \underline{0.921} & 0.625 & 0.746 & 0.516 & 0.710 \\
Mistral 7B-Inst             & \underline{0.821} & 0.918 & 0.692 & 0.786 & 0.547 & 0.770 \\
\bottomrule
\end{tabular*}

    \label{tab:e2e_performance}
\end{table}

Table \ref{tab:e2e_performance} demonstrates that our end-to-end linking approach outperforms baseline methods across all datasets. Using $\text{BM25}_3$ blocking and GPT-4o as candidate selector, we achieve F1-scores of 79-85\%. With Mistral 7B-Instruct, our best-performing local candidate selector, F1-scores range from 55-82\%. Detailed results for all models and error type analysis are available in \ref{appendix:e2e}.

\emph{This experiment shows that our approach outperforms baseline methods on all three datasets by up to 37.4\% in F1-score. Furthermore, it demonstrates the generalization capabilities of our approach by achieving F1-scores of over 79\% on all datasets.}

\section{Discussion and Conclusions}

In this paper, we addressed the issue of \emph{attribution tag quality}, with a particular focus on data inconsistencies that arise when attribution tags are shared among different parties. We argue that data quality issues can mislead forensic investigations and even result in false convictions if addresses are labeled incorrectly. To solve this, we proposed a novel computational approach based on Large Language Models (LLMs) that automatically links attribution tags to well-defined concepts in knowledge graphs, addressing the semantic nature of the problem. We implemented our approach in an end-to-end pipeline and demonstrated that, when combined with filtering and blocking techniques, it outperforms existing methods. Additionally, we showed that pre-trained LLMs running locally on consumer-grade hardware achieve performance comparable to remote models. Furthermore, we demonstrated that carefully designed prompts can significantly reduce costs with only a marginal decrease in performance. Overall, we believe our approach not only addresses the pressing issue of inconsistent attribution tags but also has the potential to inspire broader efforts to improve data quality in other forensic investigation tools and platforms.

One limitation of our approach is its binding to specific application domains; so far, we lack evidence that our method is generally applicable to all record linkage problems. However, we believe that a data- and measurement-driven approach would be valuable for assessing the broader suitability of this method. Another limitation is the assumption that different parties (e.g., exchanges, investigators) use the same knowledge graph when exchanging attribution tag records. If this is not the case, our approach does not harmonize the data but merely shifts the problem to a different abstraction level. However, significant efforts are being made to harmonize and adopt shared knowledge graphs within the field. For example, the Darkweb and Virtual Assets taxonomy developed by INTERPOL has been integrated into the Malware Information Sharing Platform (MISP) Galaxy\footnote{\url{https://misp-galaxy.org/interpol-dwva/}}. This taxonomy helps categorize and enrich threat intelligence, with MISP Galaxies organizing related data clusters to describe higher-level concepts such as adversary groups, malware families, and vulnerabilities, simplifying complex data analysis for organizations.

Future work could include fine-tuning pre-trained LLMs to the cryptoasset domain, improving performance by recognizing semantically related terms that are syntactically different. Additionally, leveraging relationships like hypernyms and hyponyms in knowledge graphs could refine candidate matching. Extending the approach to automatically categorize cryptoasset addresses based on well-defined categories would also provide a more comprehensive solution.

\begin{credits}
\subsubsection{\ackname} Avice gratefully acknowledges funding from the Digital Finance CRC which is supported by the Cooperative Research Centres program, an Australian Government initiative. This research was partially funded by the Austrian security research program KIRAS of the Federal Ministry of Finance (BMF) under the project DeFiTrace (grant agreement number 905300) and the FFG BRIDGE project AMALFI (grant agreement number 898883).

\subsubsection{\discintname}
The authors have no competing interests to declare that are relevant to the content of this article.
\end{credits}

\clearpage
\begin{appendix}
\renewcommand{\thesection}{\appendixname~\arabic{section}}
\section{}  \label{appendix:cg}
\begin{figure}
    \centering
    \includegraphics[width=\columnwidth]{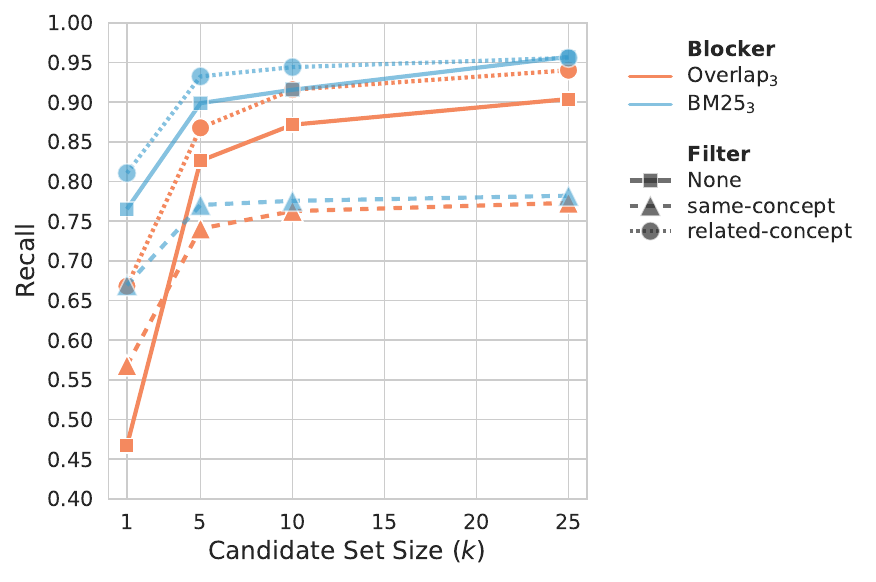}
    \caption{\textbf{Choosing Candidate Set Size}. Recall of candidate set generation for different candidate set sizes ($k$). The performance gain for all methods visibly slows down for $k > 5$.}
    \label{fig:cg_results}
\end{figure}

A visual illustration of the candidate set generation performances of the different blocking and filtering techniques can be found in Figure~\ref{fig:cg_results}. We can see that the recall gain from $k=1$ to $k=5$ is significant, while subsequent increases in candidate set size have only a marginal impact.

\section{} \label{appendix:el}
\begin{figure}
    \centering
    \includegraphics[width=\textwidth]{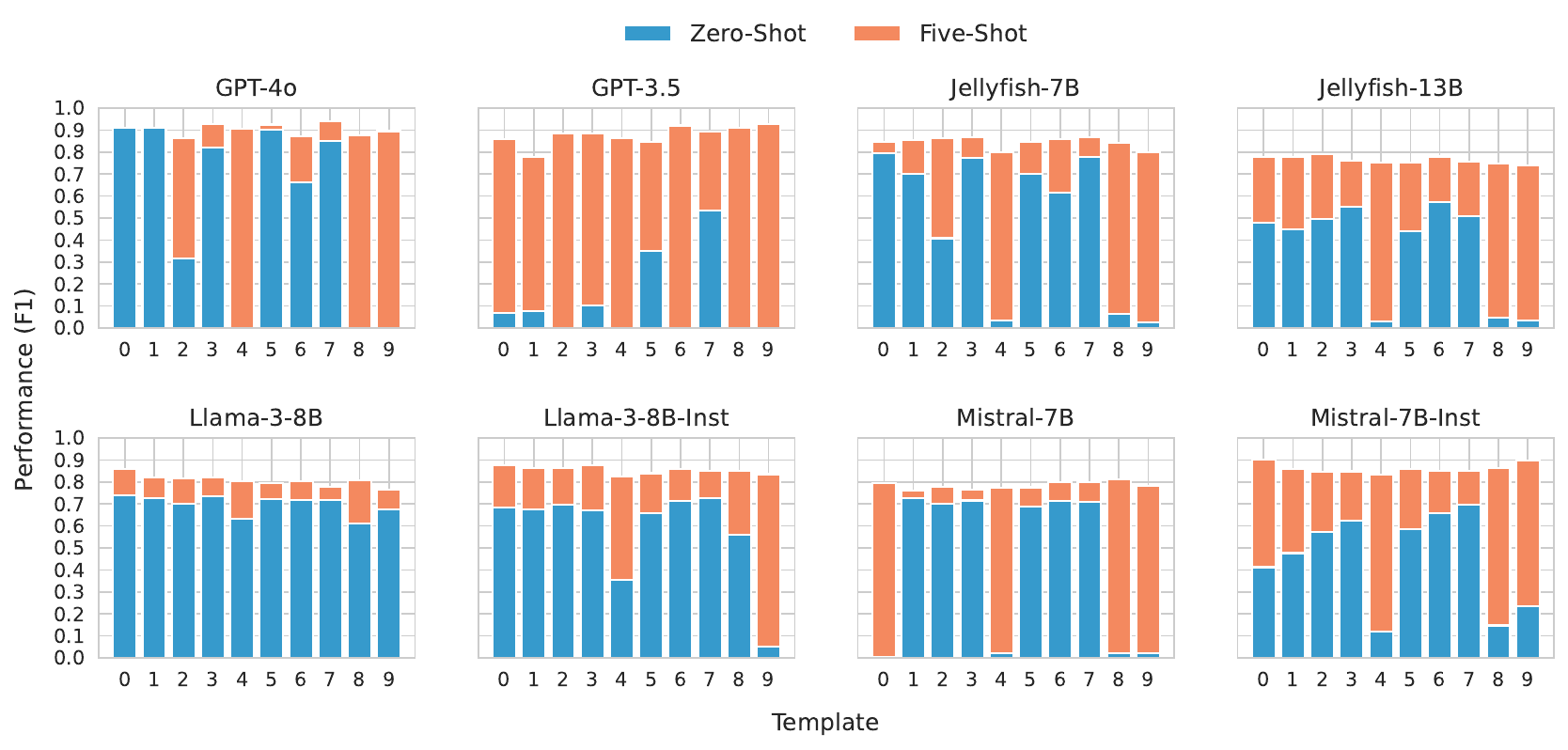}
    \caption{\emph{Candidate Selection Experiment:} Model performances ($\mathbf{F1}$) across each template using no (\emph{Zero-Shot}) and five (\emph{Five-Shot}) examples in the prompt.}
    \label{fig:cs_result_full}
\end{figure}

A comprehensive overview of all candidate selector model results across each template can be found in Figure~\ref{fig:cs_result_full}. We can see that the zero-shot performance of the models varies significantly across templates; however, introducing five examples reduces this variance considerably. Interestingly, while GPT-3.5 is ineffective without examples, with five examples it achieves the second-highest F1-score. Also notable is that the zero-shot performance for templates 4, 8, and 9 is lowest, signaling that the output format reminder (OUT) is crucial for good results without examples.

\section{} \label{appendix:e2e}

In our experiments, we treat entity linking as a multi-classification problem and measure the performance with the macro F1 score and the accuracy, see Table~\ref{tab:e2e_performance_full}. For the macro F1-score, this means that we calculate each entity's individual F1 score and average them. Consequently, any prediction that is different from the ground truth is treated equally as an error. In a practical setting, we argue that errors predicting no entity matches are less problematic than errors linking the attribution tag to the wrong entity.

\begin{table}
    \centering
    \caption{Full overview of end-to-end entity linking performance including the results of all models the three datasets}
    \begin{tabular*}{\textwidth}{@{\extracolsep{\fill}}lccccccc}
\toprule
\multirow{2}{*}{\textbf{Model}} & \multicolumn{2}{c}{\textbf{GraphSense}} & \multicolumn{2}{c}{\textbf{WatchYourBack}} & \multicolumn{2}{c}{\textbf{DeFi Rekt}} \\
\cmidrule(lr){2-3} \cmidrule(lr){4-5} \cmidrule(lr){6-7}
 & \textbf{F1} & \textbf{Acc.} & \textbf{F1} & \textbf{Acc}. & \textbf{F1} & \textbf{Acc.} \\
\midrule
$\text{BM25}_3$             & 0.718 & 0.789 & 0.495 & 0.516 & 0.393 & 0.510 \\
UnicornPlus                 & 0.667 & 0.445 & 0.558 & 0.651 & 0.352 & 0.670 \\
UnicornPlusFT               & 0.783 & 0.873 & 0.542 & 0.603 & 0.419 & 0.610 \\
\\
GPT4o                       & \textbf{0.853} & \textbf{0.927} & \textbf{0.801} & \textbf{0.873} & \textbf{0.793} & \textbf{0.930} \\
GPT3.5                      & 0.810 & 0.881 & \underline{0.756} & \underline{0.825} & \underline{0.691} & \underline{0.890} \\
Jellyfish 7B                & 0.799 & 0.914 & 0.634 & 0.746 & 0.432 & 0.640 \\
Jellyfish 13B               & 0.716 & 0.842 & 0.595 & 0.730 & 0.478 & 0.720 \\
Llama 3 8B                  & 0.785 & 0.917 & 0.618 & 0.683 & 0.440 & 0.680 \\
Llama 3 8B-Inst             & 0.814 & \underline{0.921} & 0.625 & 0.746 & 0.516 & 0.710 \\
Mistral 7B                  & 0.745 & 0.887 & 0.467 & 0.452 & 0.327 & 0.330 \\
Mistral 7B-Inst             & \underline{0.821} & 0.918 & 0.692 & 0.786 & 0.547 & 0.770 \\
\bottomrule
\end{tabular*}
    \label{tab:e2e_performance_full}
\end{table}

\begin{figure}
    \centering
    \includegraphics[width=\textwidth]{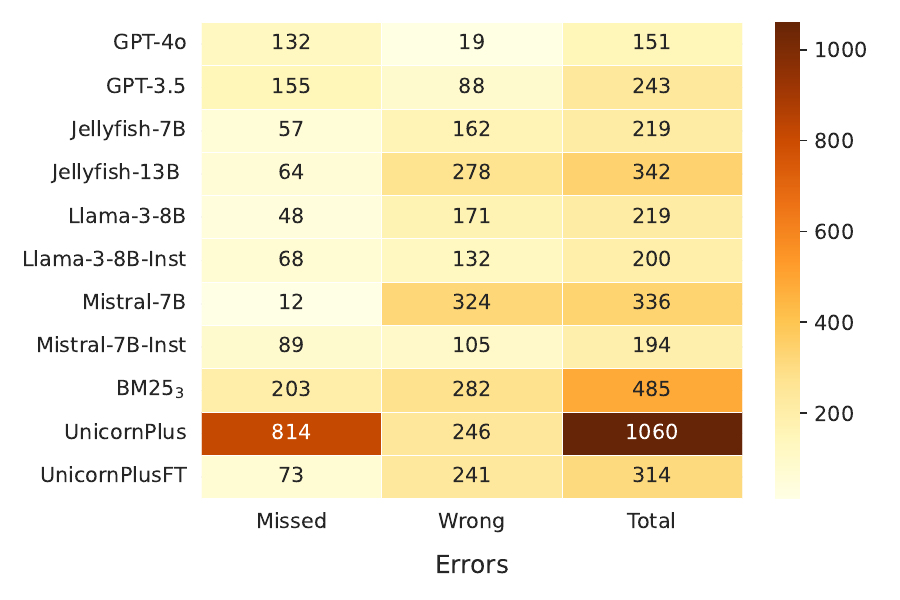}
    \caption{\textbf{Error analysis.} The number of missed and wrongly predicted actor links for each model in the end-to-end entity linking experiment. The experiment contained 2000 samples from three different datasets.}
    \label{fig:error}
\end{figure}

Figure \ref{fig:error} shows us the error composition of the different models across all three datasets. We distinguish between the \emph{Missed Entity} error from a wrong no-match prediction and the \emph{Wrong Entity} error with a wrong entity predicted. We can see that GPT-4o not only has the best performance in terms of accuracy and macro F1-score but also has less than 1\% of wrong entities predicted, with the majority of the errors coming from missed entities. In contrast, the local LLMs all have more \emph{Wrong Entity} than \emph{Missed Entity} errors. Our best-performing local LLM, Mistral 7B-Instruct, has, despite better overall accuracy, more \emph{Wrong Entity} errors than GPT-3.5.

In Table~\ref{tab:e2e_wrong_actors}, we identify the three actors most frequently misclassified by each model. The results are similar across different models, with a few exceptions. For example, \texttt{maker} appears in the top three only for UnicornPlus, despite being present in 135 different tags. Notably, none of the models correctly classify any of the 20 tags associated with \texttt{0x} as an actor.

\begin{table}
    \centering
    \caption{Overview of the most frequently misclassified actors for each model}
    \begin{tabular*}{\textwidth}{@{\extracolsep{\fill}}llrr}
\toprule
Model & Actor & \# Miss Classified & \# Total Occurences \\
\midrule
\multirow{3}{*}{$\text{BM25}_{3}$} & aave & 121 & 133\\
                                   & curve & 38 & 128 \\
                                   & binance & 33 & 37\\
\addlinespace[0.5em]                                   
\multirow{3}{*}{UnicornPlus} & aave & 133 & 133 \\
                             & maker & 128 & 135\\
                             & synthetix & 122 & 198\\
\addlinespace[0.5em]
\multirow{3}{*}{UnicornPlusFT} & huobi & 51 & 53\\
                               & aave & 35 & 133\\
                               & 0x & 20 & 20\\
\addlinespace[0.5em]
\multirow{3}{*}{GPT4o}  & 0x & 20 & 20\\
                        & aave & 18 & 133\\
                        & feiprotocol & 10 & 27\\
\addlinespace[0.5em]                        
\multirow{3}{*}{GPT3.5} & synthetix & 30 & 198\\
                        & aave & 22 & 133\\
                        & 0x & 20 & 20\\
\addlinespace[0.5em]                        
\multirow{3}{*}{Jellyfish-7B} & aave & 29 & 133\\
                              & 0x & 20 & 20\\
                              & feiprotocol & 10 & 27\\
\addlinespace[0.5em]
\multirow{3}{*}{Jellyfish-13B} & aave & 52 & 133\\
                               & 0x & 20 & 20\\
                               & synthetix & 19 & 198\\
\addlinespace[0.5em]                               
\multirow{3}{*}{Llama 3 8B} & 0x & 20 & 20\\
                            & aave & 20 & 133\\
                            & feiprotocol & 10 & 27\\
\addlinespace[0.5em]                           
\multirow{3}{*}{Llama 3 8B-Inst } & aave & 24 & 133\\
                                  & 0x & 20 & 20\\
                                  & feiprotocol & 11 & 27\\
\addlinespace[0.5em]                             
\multirow{3}{*}{Mistral 7B}  & aave & 23 & 133\\
                             & 0x & 20 & 20\\
                             & curve & 19 & 128\\
\addlinespace[0.5em]
\multirow{3}{*}{Mistral 7B-Inst} & aave & 21 & 133\\
                                 & 0x & 20 & 20\\
                                 & feiprotocol & 10 & 27\\

\bottomrule
\end{tabular*}
    \label{tab:e2e_wrong_actors}
\end{table}

\end{appendix}

\clearpage

\bibliographystyle{splncs04}
\bibliography{references}

\end{document}